\documentclass[10pt, conference, compsocconf]{IEEEtran}

\ifCLASSOPTIONcompsoc
  \usepackage[nocompress]{cite}
\else
  \usepackage{cite}
\fi

\ifCLASSINFOpdf
  \usepackage[pdftex]{graphicx}
\else
\fi

\usepackage{url}
\usepackage{color,soul}
\usepackage{amsmath}
\usepackage{siunitx}

\begin{document}

\title{EscapeWildFire: Assisting People to Escape Wildfires in Real-Time}



%
\author{\IEEEauthorblockN{Andreas Kamilaris\IEEEauthorrefmark{1}\IEEEauthorrefmark{2},
Jesper Provoost\IEEEauthorrefmark{2},
Jean-Baptiste Filippi\IEEEauthorrefmark{3}, 
Chirag Padubidri\IEEEauthorrefmark{1}, 
Savvas Karatsiolis\IEEEauthorrefmark{1},\\
Ian Cole\IEEEauthorrefmark{1}, 
Wouter Couwenbergh\IEEEauthorrefmark{2} and
Evi Demetriou\IEEEauthorrefmark{1}}
\IEEEauthorblockA{\IEEEauthorrefmark{1}CYENS Center of Excellence, Nicosia, Cyprus\\
Email: \{a.kamilaris, c.padubidri, s.karatsiolis, i.cole, e.demetriou\}@cyens.org.cy }
\IEEEauthorblockA{\IEEEauthorrefmark{2}Department of Computer Science, University of Twente, Enschede, The Netherlands\\
Email: \{j.c.provoost, w.couwenbergh\}@student.utwente.nl}
\IEEEauthorblockA{\IEEEauthorrefmark{3}Centre national de la recherche scientifique (CNRS) SPE, Università di Corsica, France\\
Email: filippi\_j@univ-corse.fr}}

\maketitle

\begin{abstract}
Over the past couple of decades, the number of wildfires and area of land burned around the world has been steadily increasing, partly due to climatic changes and global warming. Therefore, there is a high probability that more people will be exposed to and endangered by forest fires. Hence there is an urgent need to design pervasive systems that effectively assist people and guide them to safety during wildfires. This paper presents EscapeWildFire, a mobile application connected to a backend system which models and predicts wildfire geographical progression, assisting citizens to escape wildfires in real-time. A small pilot indicates the correctness of the system. The code is open-source; fire authorities around the world are encouraged to adopt this approach.
\end{abstract}


%
\IEEEpeerreviewmaketitle

\section{Introduction}
Over the past couple of decades the number of wildfires and area of land burned has been steadily increasing  \cite{dennison2014large}, partly due to climatic changes and global warming \cite{abatzoglou2016impact}. A wildfire is defined as a large, destructive fire that spreads quickly over woodland or brush. It differs from fires that occur in urban infrastructures and buildings. Every year approximately half a million hectares of land are burned by wildfires in southern Europe \cite{moreira2011landscape}. Moreover, the frequency of these fires is projected to increase by 27\% in the coming decade \cite{huang2015sensitivity, forzieri2016multi}. There is a high chance that more people will be endangered by wildfires in the coming years and this will likely be associated with further casualties \cite{forzieri2016multi}. 
Countries tend to deal with increasing wildfire hazards by improving their equipment and capacity in human resources, employing emerging technologies such as drones \cite{KamilarisSept2017aDisMan, kamilaris2019training}. However, the infrastructures of most countries are lacking in the evacuation of humans from the fire zones, despite evacuation plans being in place. Furthermore, the evacuation plans in regions where forest fires are not common (e.g. northern Europe such as Netherlands and Nordic countries) have not been fully implemented or tested for safety \cite{kuligowski2020evacuation}.

The contribution of this paper is to address this gap and help tackle the limited preparedness of many countries and regions around the world for wildfire evacuation. A mobile application and backend, which assist people in danger to safely escape wildfires in real-time, are presented in this work. This software has been created as open-source code\footnote{EscapeWildFires software. https://github.com/rise-centre/Escape-Wildfires}, together with complete user manuals and instructions. The authors wish to encourage fire authorities around the world to embrace and implement this approach.

The rest of the paper is organized as follows: Section \ref{RelWork} presents related work and Section \ref{Method} describes design and development of the EscapeWildFire system. Then, Section \ref{evaluation} explains our evaluation efforts through a case study at the region of Nicosia, Cyprus while Section \ref{Results} shows the evaluation results. Finally, Section \ref{Discussion} discusses overall implications and Section \ref{Conclusion} concludes this paper.

\section{Related Work}
\label{RelWork}
Related work in the field of fire evacuation has mainly focused on evacuation from inside buildings \cite{shi2009agent}, predominantly public buildings. Some authors have worked on evacuating neighborhoods \cite{cova2002microsimulation}, trains \cite{capote2012analysis}, hospitals \cite{sternberg2004counting} and road tunnels \cite{nilsson2009evacuation}. Authors have investigated different protocols, escape routes, time required to evacuate depending on building's floor or position inside the road tunnel, human behavioral patterns during fires, congestion, etc. Some evacuation models have been used to assess and analyze the safety of various infrastructures \cite{nilsson2009evacuation, capote2012analysis} and to forecast and visualize the wildfire in geo-visualizations \cite{castrillon2011forecasting}.

Regarding wildfires in particular, related scientific work targeting real-time evacuation assistance is very limited. WFA Pocket \cite{monedero2019predicting} is a tool targeted for fire fighters, designed to be used in the field, modelling the progression of the fire based on the user inputs, such as the fuel types, real-time weather data, etc.
WiSE\footnote{Wildfire Safety Evaluator, https://wise.wildfireanalyst.com} is a tool designed to provide safe separation distance (SSD) calculations between wildfires and fire fighters.
Gaia GPS\footnote{Gaia GPS, https://www.gaiagps.com} is a mobile app that has been used by firefighters to plan escape routes, mark fire lines, and track progression. Finally, Emergency is an application developed by the American Red Cross \cite{bachmann2015emergency}, with the aim to keep people safe in severe weather conditions as well as man-made or natural disasters. 

\section{Design and Development}
\label{Method}
This section discusses the design decisions and implementation details of the fire evacuation software.

\subsection{Mobile App Design}
\label{mobile_app}
\begin{figure}[!t]
\centering
\includegraphics[width=\linewidth]{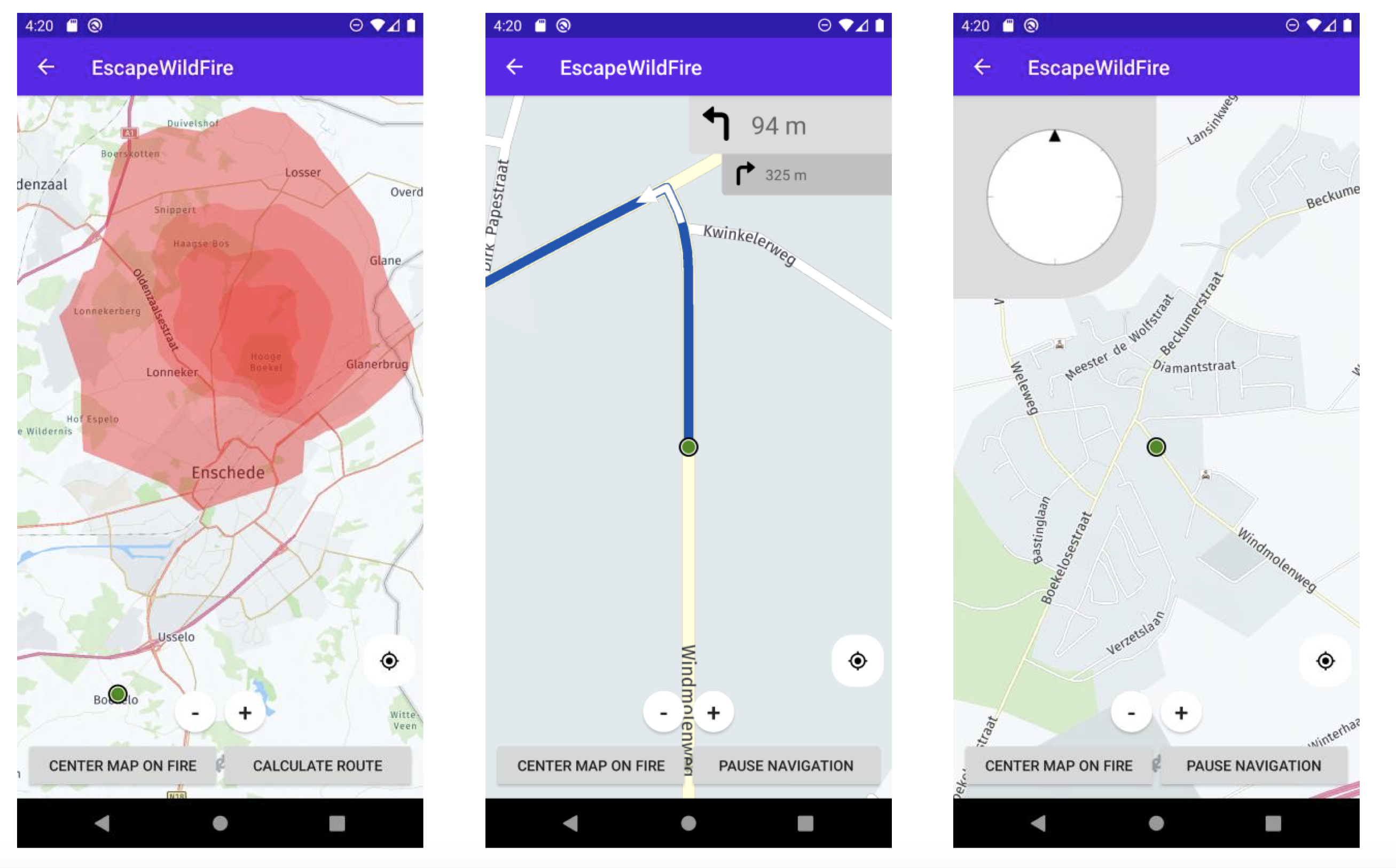}
\vspace{-0.5cm}
\caption{Snapshots of the mobile app: Fire illustration (left), turn-by-turn view (center), direction-based view (right).}
\label{snapshots24}
\end{figure}

The main idea is to maintain a minimalist design approach (see Figure \ref{snapshots24}). 
The fire location and progression indicator is red, with dark red at the center of the fire and increasing transparency further from the center, indicating time periods of fire progression (Figure \ref{snapshots24}, left). Each polygon of increasing transparency indicates the expected propagation of wildfire after 15 minutes. For example, in Figure \ref{snapshots24} (left), there are five layers of the wildfire, at 0, 15, 30, 45 and 60 minutes after ignition. 
Two different navigational methods were considered: a \textit{turn-by-turn} view (similar to that used in Google Maps) and a \textit{direction-based} view (similar to compass navigation).
For the turn-by-turn view, (Figure \ref{snapshots24}, center),the top-right corner indicates the next turn and the turn immediately after, along with the distance until action. At the bottom of the screen, the user has the option to open a complete list of directions. This is a scroll-enabled list, listing turn types alongside the distance to each turn. Next to the direction list is another button to pause navigation, returning the user's screen to an overview of the route. Above these buttons, two zoom buttons are placed, offering the option for fine zoom control.
For the direction-based view, an arrow inside a white circle (Figure \ref{snapshots24}, right) continuously shows the indicated direction of the user, based on current location and orientation.

\subsection{Backend Implementation}
\begin{figure*}[!t]
\centering
\includegraphics[width=0.9\linewidth]{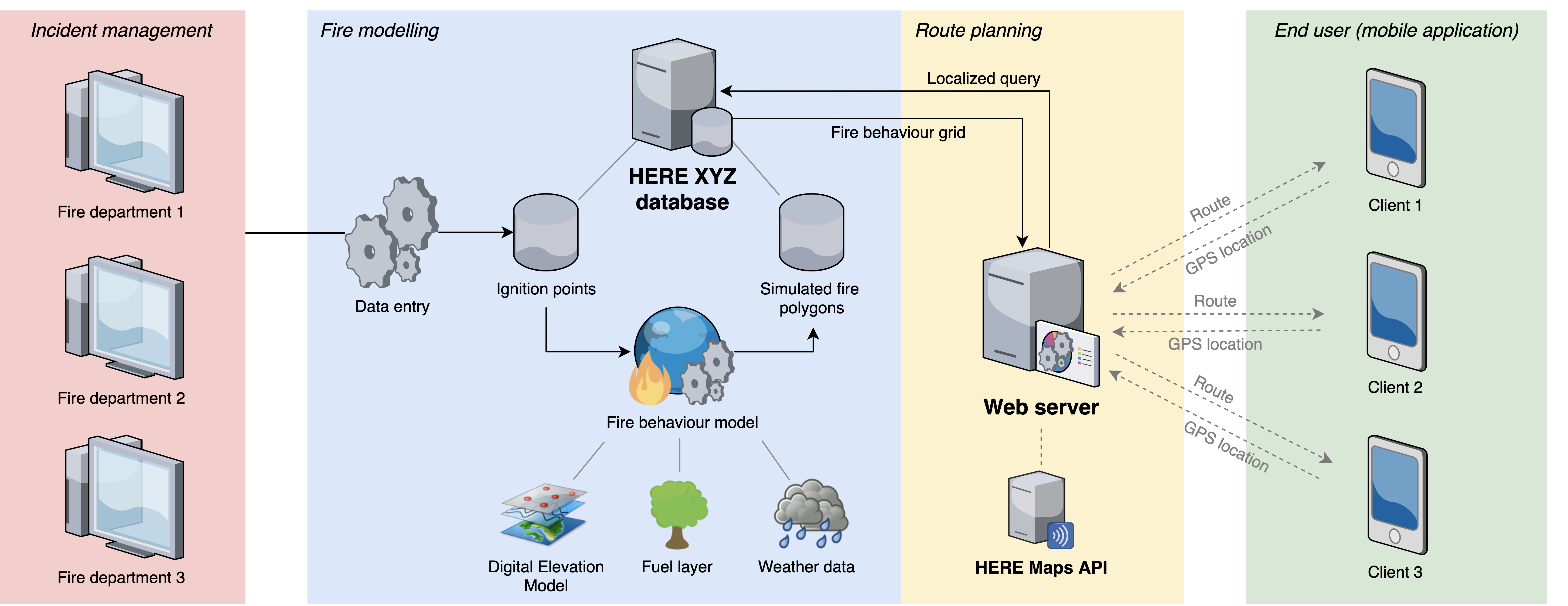}
\vspace{-0.2cm}
\caption{Architecture of the EscapeWildFire system.}
\label{framework}
\end{figure*}

Besides the mobile app, a backend system supports the management of vital information, such as where/when the fire started, how fast it will propagate based on vegetation (fuel) type, wind speed and direction, weather conditions, etc. The architecture of the backend, in relation to the mobile app, is provided in Figure \ref{framework}. The architecture is composed of the following components:
\begin{itemize}
    \item Mobile application (see Sections \ref{mobile_app} and \ref{Android_app})
    \item Fire behaviour model (see Section \ref{fire_behaviour})
    \item Web server (see Section \ref{Web_server})
    \item Fire management tool (see Section \ref{man_software})
\end{itemize}

Each fire department has its own version of the fire management tool (FMT), which allows them to add ignition points in areas of their responsibility. The fire behaviour model (FBM) analyzes the area around the ignition points and calculates the fire propagation. The Web server communicates the information from the FBM to the mobile apps of the end users and data is presented in the form of polygons, indicating the spread of the wildfire in 15-minute intervals, up to 60 minutes ahead of time. The mobile app calculate safe escape routes for the users for effective evacuation.

\subsubsection{Mobile application}
\label{Android_app}
Implemented in Android, based on the design decisions listed in Section \ref{mobile_app}. The map provider is HERE maps\footnote{HERE Maps. https://mobile.here.com/?x=ep}. The app has been listed on Google Play\footnote{Escape Wildfire mobile app. https://play.google.com/store/apps/details?\\id=com.ewf.escapewildfire} to facilitate the evaluation process (i.e., for pilot participants to easily download, see Section \ref{Results}). The most important algorithm is the one used for calculating escape routes. 
Each possible route gets a score, depending on the time needed to reach \textit{safety}, where safety is defined as a one kilometer (km) distance from the wildfire after one hour from its ignition time, plus the angle from the fire's propagation direction:
\begin{equation}
    \stackrel{*}{r}  = argmax \ \{  r(1), r(2), ... , r(x) \}  \\~\\ 
\end{equation}
where $\stackrel{*}{r}$ the best route selected, $x$ the total number of possible routes and each route is calculated as:

\begin{equation}
\begin{aligned}
    r(i) = a(i) \times \textit{angle}(i)+(1-a(i)) \times \frac{1}{\textit{time\_to\_safety}(i)} \textit{,} \\ 0<a(i)<1
\end{aligned}
\end{equation}

where $angle$ is the angle between direction of wildfire and suggested route, $time\_to\_safety$ is the time needed to reach a safe place, while the parameter $\alpha$ suggests the weight of each parameter, which differs between transport types (matter of future work).
Routes that pass through wildfire zones (see Figures \ref{snapshots24} and \ref{region1}) during the 15-minute slots when the wildfire is expected to reach those zones are rejected immediately. Travel calculations are considered based on conservative values for user travel speed (Walking: 4.5 km/h, Cycling: 15 km/h, Driving car: 50 km/h).

\subsubsection{Fire behaviour model}
\label{fire_behaviour}
Wildland fire spread simulations are performed by the numerical solver ForeFire~\cite{filippi10discrete}. ForeFire relies on a front-tracking method where the fire front is represented by Lagrangian markers that are linked to each other via a dynamic mesh. While the tool can theoretically use any formulations, currently the velocity of every point of the front is provided by the model of Rothermel~\cite{rothermel72mathematical}. The rate of spread (ROS) is expressed as a function of several environmental properties such as wind speed, terrain slope, fuel moisture content and other fuel parameters characterizing the vegetation. A simulation mostly consists of the definition of an initial state of the fire front and then the ROS is computed for the markers of the fire front based on underlying 2D fields from which environmental properties are determined. ForeFire relies on a discrete-event approach where most computations deal with the determination of the time at which the markers will reach their next destination. Destination here is defined by a fixed spatial increment in the direction at the tangent to the perimeter of the burned area. Real-time wind information is provided by the Windy\footnote{Windy. https://www.windy.com} online service.
The land type distribution field comes from CORINE Land Cover data~\cite{CLC}. The elevation field is extracted from the NASA SRTM dataset, which originally has a 30 m resolution. Fuel parametrization is performed to assign reference fuel parameters to each type of vegetation in the land use data for the ROS computations. Data includes 2D fields of wind speed vectors, to account for the influence of the elevation field.

\subsubsection{Web server}
\label{Web_server}
The Web server is responsible for linking the mobile apps of the end users with the backend (i.e. FMT and FBM). The Web server uses the HERE HYZ database and its Maps API in order to properly model fire propagation as polygons, in a language understood by the map provider software used by the mobile app. HERE HYZ\footnote{HERE Studio. https://developer.here.com/products/platform/studio} is a location data management cloud service, built around standards like GeoJSON. Each fire is modelled as a GeoJSON geometry object. When accessing the service through HERE Maps API, communication is through GeoJSON files.

\subsubsection{Fire management tool}
\label{man_software}
\begin{figure*}[!t]
\centering
\includegraphics[width=0.8\linewidth]{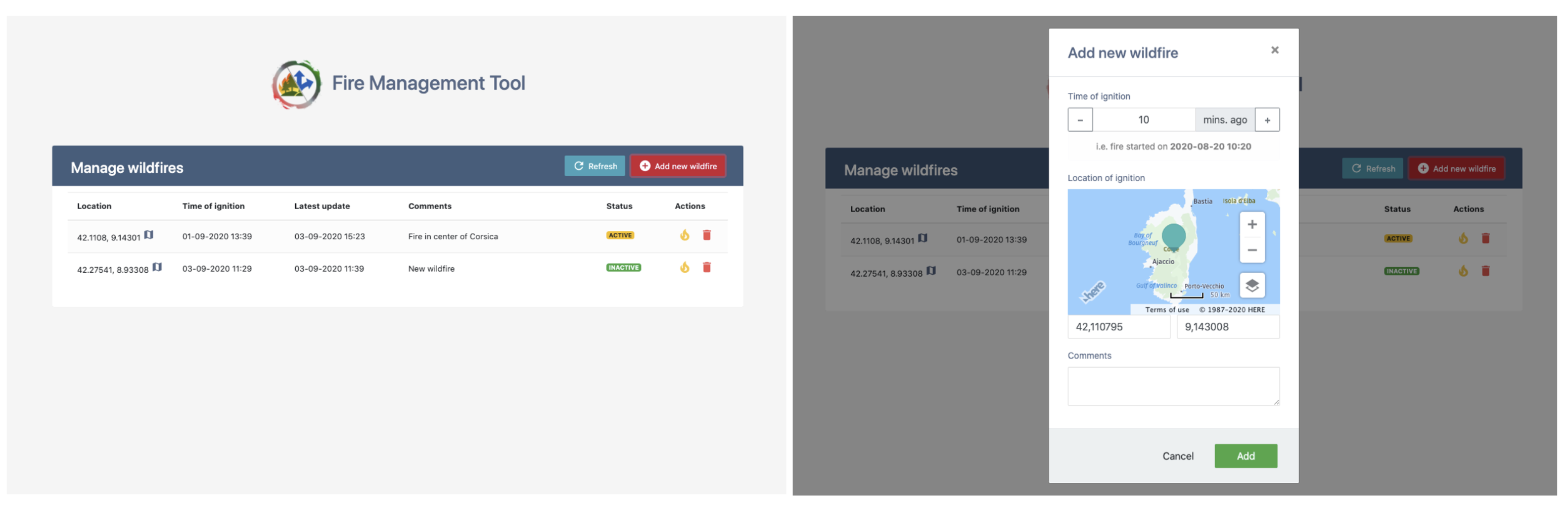}
\vspace{-0.3cm}
\caption{Snapshots of the fire management tool.}
\label{incidentsoft}
\end{figure*}

This software is intended to be used by fire departments to manage wildfire incidences, i.e. add/remove, ignite/stop and configure wildfires in real-time. New fires can be added via the \textit{"Add new wildfire"} button, shown in Figure \ref{incidentsoft} (left). When the fire officer clicks this button, relevant details can be added (Figure \ref{incidentsoft}, right) such as the exact location where the wildfire started, time of ignition and other important information to be shared with stakeholders (i.e. fire fighters and citizens).

\section{Evaluation}
\label{evaluation}
\begin{figure*}[!t]
\centering
\includegraphics[width=0.8\linewidth]{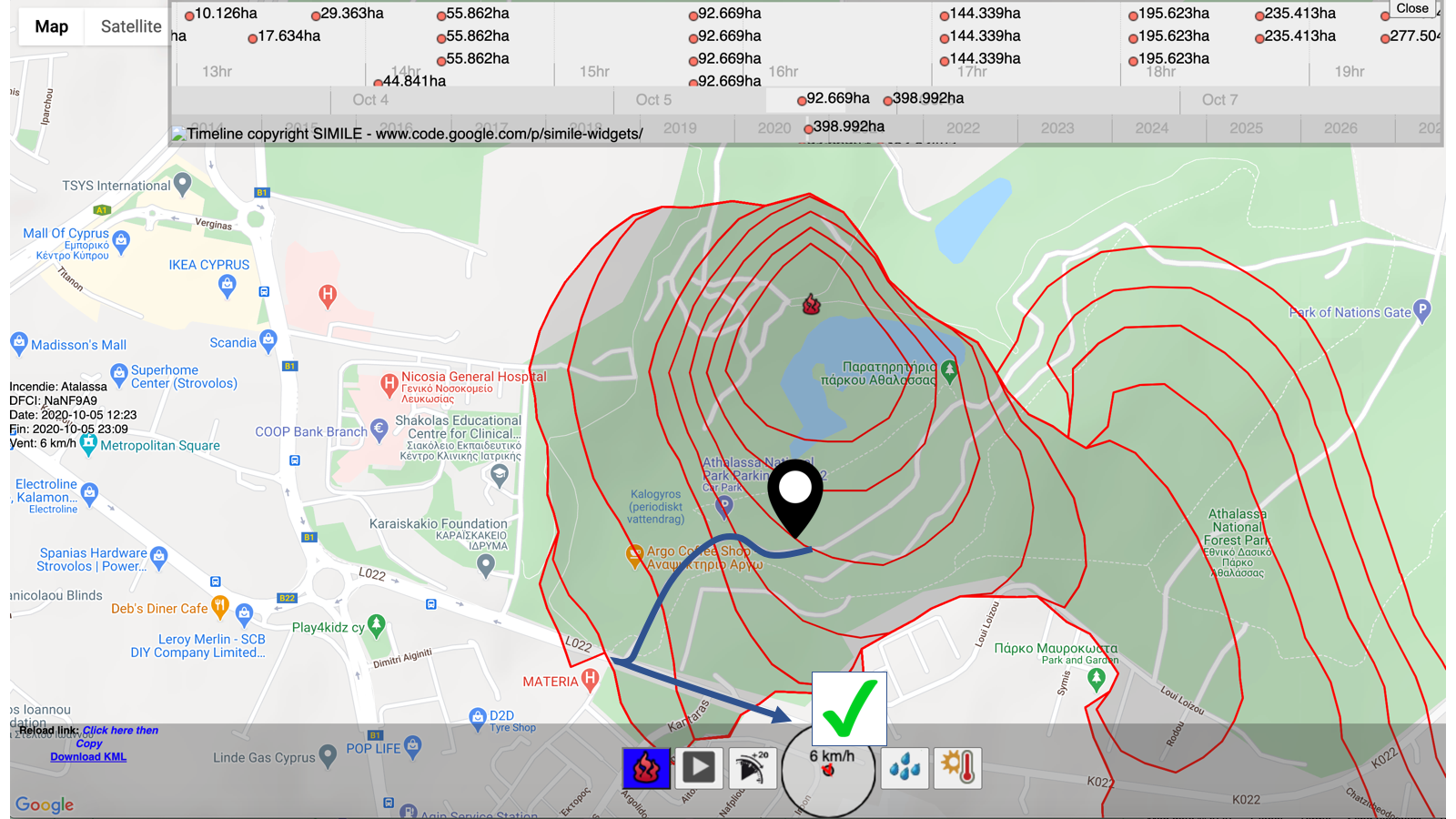}
\vspace{-0.2cm}
\caption{Evaluation area: Athalassa natural reserve. The ignition of wildfire, participants' initial positions and suggested escape route.}
\label{region1}
\end{figure*}

To assess the proposed system, a scenario of a wildfire was simulated selecting a natural reserve as the landscape type. This landscape is located in the region of Nicosia, Cyprus, illustrated in Figure \ref{region1}, as visualized by the ForeFire fire behaviour modeller \cite{filippi10discrete} (see Section \ref{fire_behaviour}). In the figure, fire has been ignited near the center of the map, while red circles show the spread of the fire in subsequent one-hour slots. At the top of the window, the hectares covered by the fire at each hour after ignition are displayed, while at the bottom, parameters such as wind speed, humidity and temperature can be set. Table \ref{table_params} shows the characteristics of the scenario under study.

\begin{table*}[!t]
\caption{Scenario under study and characteristics/parameters.}
\label{table_params}
\centering
\begin{tabular}{|p{1.8cm}|c|c|p{1.9cm}|c|p{1.1cm}|c|p{2.5cm}|}
\hline
\textbf{Area} & \textbf{Type} & \textbf{Participants} & \textbf{Transport} & \textbf{Wind} & \textbf{Humidity} & \textbf{Temperature} & \textbf{Time to evacuate} \\
\hline
Athalassa & Natural reserve & 17 & Walking & 6 km/h (east-south) & 30\% & \ang{30} Celsius & 20 minutes \\
\hline
\end{tabular}
\end{table*}

Participants were selected randomly as they were passing by on foot. They were asked to participate in our pilot study simulating the scenario of a fire occurring near them. The evaluation exercise took place in the afternoon (around 17:00-19:00), coinciding with a high likelihood of free time. Participants were asked to download the mobile app from Google Play and then follow the instructions given by the app. Afterwards, they were asked to fill a questionnaire, eliciting information about demographics, plus the questions listed in the first two columns of Table \ref{questionnaire}. As compensation for their efforts, participants received a voucher for free coffee at a popular café in the local area.

\begin{table*}[!t]
\caption{Questionnaire used during the evaluation process.}
\label{questionnaire}
\centering
\begin{tabular}{|p{6.3cm}|p{4.5cm}|p{4.1cm}|}
\hline
\textbf{Question} & \textbf{Type} & \textbf{Responses} \\
\hline
How easy was the application to understand in general? & Likert scale, from very difficult (1) to very easy (5) & Avg. value: 3.9 St. dev.: 0.9  \\
\hline
How easy was the application to follow using the advice given? & Likert scale, from very difficult (1) to very easy (5) & Avg. value: 3.6 St. dev.: 0.8   \\
\hline
Did you feel safer while using the application, although there was a dangerous situation near you? & Yes/No & Yes (82\%), No (18\%) \\
\hline
Would you keep this app downloaded on your phone? & Yes/No & Yes (94\%), No (6\%)  \\
\hline
Did the application help you escape the fire safely? & Yes/No & Yes (100\%)  \\
\hline
Did you feel like the application worked the way it should have? & Yes/No & Yes (100\%)  \\
\hline
How likely is it to suggest this application to someone? & Likert scale, from very unlikely (1) to very likely (5) & Avg. value: 4.1 St. dev.: 0.7   \\
\hline
How likely is it to use this application in a real life scenario? & Likert scale, from very unlikely (1) to very likely (5) & Avg. value: 4.1 St. dev.: 1.2   \\
\hline
What improvements would make the app more useful, effective and/or user friendly? & Open question & See Section \ref{Discussion}. \\
\hline
How much time did you need to escape the fire? & Response in minutes:seconds & (for 0.8 km) Avg. value: 8:30 St. dev.: 5:25  \\
\hline
Did you follow 100\% the route suggested by the app? If not, why? & Open question & Yes (94\%), No (6\%) \\
\hline
Was the route suggested by the app the best one to your knowledge, in order to safely evacuate? & Open question & Yes (100\%) \\
\hline
\end{tabular}
\end{table*}

\section{Results}
\label{Results}
Seventeen people in total agreed to participate in the pilot. Their demographic information is shown in Table \ref{table_demographics}. Gender was equally balanced. The large majority of participants were locals. Most participants were well educated, possessing as diploma, bachelor or master level degree. 
The last three columns of Table \ref{questionnaire} show the responses of the participants after using the mobile app to safely escape the fire. The large majority found the application easy to use, with little effort required to follow the advice given. The vast majority felt they would be safer using the app when in actual danger and would keep it downloaded on their mobile phones. Two participants (females, 43 and 67 years) argued that during a fire they cannot feel safe with or without an mobile app as assistance.

\begin{table*}[!t]
\caption{Demographic info of participants.}
\label{table_demographics}
\centering
\begin{tabular}{|p{1.1cm}|p{2.0cm}|p{2.2cm}|p{2.5cm}|p{3.0cm}|p{3.0cm}|}
\hline
\textbf{Area} & \textbf{Participants} & \textbf{Gender} & \textbf{Ages} & \textbf{Nationality} & \textbf{Education}  \\
\hline
Athalassa & 17 & Male (8), Female (9)  & 16-20 (1), 21-35 (5), 36-50 (6), 51+ (5) & Cypriot (14), Greek (2), Kurdish (1) & Diploma (5), Bachelor (10), Master (2) \\
\hline
\end{tabular}
\end{table*}

In all 17 cases, the app helped the users to successfully evacuate the area. All but one participant felt like the app worked appropriately, following the route suggested by the app. All participants agreed that the route suggested by the app was, to their knowledge, the best available for safely escaping the wildfire.
To escape the fire, people needed around $8:30$ minutes to walk the $800$ meters to safety ($5.64$ km/h, see Figure \ref{region1}). There was a large standard deviation in time ($5:25$ minutes) for older people (51+) needing considerably more time to walk than younger people (16-35) ($5:50$ vs. $13:06$ minutes). This indicates that the app should consider the physical condition of the user, partly reflected through age although other metrics can be used. 

\section{Discussion}
\label{Discussion}
This paper has presented a solution to the problem of assisting people to safety during wildfires by evacuation. Various technologies have been employed to build a robust solution. The evaluation focused mostly on the feasibility of the proposed system and the correctness of the mobile app, touching upon human factors and aspects such as acceptance and usefulness. The ForeFire fire propagation prediction model has not been evaluated in this paper, but it has been extensively evaluated both in simulation \cite{filippi10discrete, allaire2020generation} and in real-world wildfire situations \cite{filippi2014evaluation, santoni2011forest}.

The system has worked correctly during the pilot. The fire management tool added new wildfires precisely, while the Web server communicated correctly with the mobile apps of the participants to show the fires and suggest escape routes. All participants agreed that the routes suggested by the app were the best available escape routes to their knowledge (see Table \ref{questionnaire}) and also agreed that the app helped them escape the fire safely. Unfortunately, the fire behaviour model, could not be evaluated in this work. However, the model has been evaluated in related work \cite{filippi10discrete}.

In terms of acceptance and usefulness, 94\% of participants would keep the app on their mobile phones, while 82\% felt safer using the app during a wildfire event. In a Likert scale of 1 (very unlikely) to 5 (very likely), participants gave a score of $4.1$ (likely) to the question of whether they would use the app in a real-life wildfire situation. It was observed that older people (51+) found it hard to trust mobile apps, especially in emergency situations.

Concerning ease of use, participants scored $3.9$ for the question whether the app was easy to understand (Likert scale, 1: very difficult to 5: very easy). Participants gave a score of $3.6$ for the question of how easy it was to follow the advice given by the app. This perhaps reveals some difficulty of navigation outside of areas with designated roads. 

The system and pilot presented in this paper had some limitations, which will be addressed in future work. More detailed land-use datasets and more specific fire velocity models need to be included, while pilots should engage more participants in a range of different landscapes and modes of transport (e.g. cycling, driving a car, etc.), considering scenarios where users do not have much time to evacuate wildfire and need to rush.
Finally, an important limitation is that our solution does not embrace or provision solutions for people with disabilities.

An important benefit of the evaluation process was the collection of useful suggestions by the participants. It was evident that older people required better accessibility features on the mobile app (e.g. larger buttons, vivid colors, flashing directions and arrows, verbal/textual directions and visual features, etc.).
There was a need to accurately define and assist the orientation of the user. This could be indicated by the phone compass showing the magnetic north whilst adjusting the map to the user orientation. Showing the historic movement of the user will also help facilitate orientation. The built-in compass of many mobile phones could also be used for areas without GPS coverage.

Some additional requirements recorded by participants in response to the question \textit{"what improvements would make the app more useful and user-friendly"} include:
\begin{itemize}
    \item \textit{Way-finding using landmarks} (e.g. mountain tops, hills, buildings), useful for some pedestrians.
    \item \textit{Assisting people in need}, i.e. locating humans who are close to the suggested escape route.
    \item Proactive information about \textit{risks in each area}.
    \item \textit{Alternative routes} calculated by monitoring users during evacuation to avoid congestion that would delay the evacuation process.
    \item A \textit{personalized speed of evacuation} for each user, depending on user's overall expected fitness, which could be partly derived from his/her age.
    \item Automatic \textit{notification for nearby wildfires} before they become life-threatening.
    \item Exploitation of \textit{civil defence infrastructures}, e.g. include routes to nearby shelters \cite{cova2009protective}.
\end{itemize}

\subsection{Future work}
An important action is to validate how the app behaves to rescue humans, assuming hypothetical placements of people near the wildfire area, examining escape paths proposed. For wildfires, we plan to consider historical data relating to  prior recordings of wildfires.
Further, future work will better consider the values of $\alpha$ and expected transport speed (see Section \ref{Android_app}), while more personalized values depending on someone's fitness or vehicle will be considered.

\section{Conclusion}
\label{Conclusion}
This paper presented EscapeWildFire, a mobile app and backend for assisting citizens to escape wildfires in real-time. The system provides a complete solution, addressing the issues of: a) recording wildfire locations, indicating the exact position and time of ignition; b) modelling and predicting the wildfire progression, based on the vegetation and fuel type around the fire, and meteorological parameters such as wind speed; and c) providing a mobile application which assists people to safety by calculating and guiding users through best evacuation routes. A small pilot study was conducted considering a natural reserve at the island of Cyprus. Results show that EscapeWildFire constitutes a correct, accepted, easy to use and realistic solution to the problem under study. Some limitations of the system and the evaluation process performed have been recorded and discussed. Various recommendations for improvements have also been recorded based on the feedback of the participants. The code of the project is available as open-source; the authors wish to encourage fire authorities around the world to adopt this approach.


%




\section*{Acknowledgements}
This project has received funding from the European Union's Horizon 2020 research and innovation programme under grant agreement No 739578 complemented by the Government of the Republic of Cyprus through the Directorate General for European Programmes, Coordination and Development.



%
\bibliographystyle{unsrt}
\bibliography{sample-base}

\end{document}